\documentstyle[prl,aps,multicol]{revtex}
\def\arx{$\alpha$--relaxation}
\def\tal{{\tau_\alpha}}

\def\A-1{\AA$^{-1}$}

\begin{document}
\draft
\title{"de Gennes" narrowing in supercooled molecular liquids :
       Evidence for center--of--mass
       dominated slow dynamics}

\author{C.~Alba-Simionesco$^1$, A.~T\"olle$^2$, D.~Morineau$^1$
        B.~Farago$^3$, G.~Coddens$^4$}
\address{$^1$Laboratoire de Chimie Physique, UMR 8611 b\^atiment 490, 
Universit{\'e} de Paris-sud, F-91405 Orsay Cedex, France}
\address{$^2$Department of Biophysical Chemistry, 
Biocenter University of Basel, CH-4056 Basel, Switzerland}
\address{$^3$Institut Laue Langevin, F-38042 Grenoble, France}
\address{$^4$Laboratoire des Solides Irradies, Ecole Polytechnique
91128, Palaiseau, France}
\date{\today}
\maketitle

\begin{abstract}
The density correlation function $\Phi(q,t)$ of the two similar
substituted aromatic liquids, Toluene and m--Toluidine, is studied by
coherent neutron spin-echo and time-of-flight scattering for wave vectors 
$q$ around the maximum $q_{\rm max}$ of the total
static structure factor $S_{\rm m}(q)$ in the supercooled region.
The wave-vector dependence of the mean structural $\alpha$-relaxation time
$\langle \tau_q \rangle$ shows in both liquids a very pronounced de 
Gennes-like narrowing centered around $q_0 < q_{\rm max}$, where $q _0$ 
corresponds to the first maximum in the center--of--mass static structure 
factor $S_{\rm COM}(q)$.
We find that the narrowing can be described {\em quantitatively} by using
$S_{\rm COM}(q)/q^2$ instead of $S_{\rm m}(q)/q^2$ indicating 
that at the corresponding molecular length scales 
the relaxation of $\Phi(q,t)$ is dominated by 
purely translational motion.
\end{abstract}
\begin{multicols}{2}
It is well known that the half width
of the experimentally observed dynamic structure factor
$S(q,\omega)$ (wave vector $q$ and frequency $\omega$)
of classical dense monoatomic fluids in equilibrium
shows a pronounced minimum at wave vectors
$q\sim q_{\rm max}\approx2\pi/\sigma$ ($\sigma$ being the atomic diameter)
where the static structure factor $S(q)$ has its first maximum.
This effect, called ``de Gennes narrowing'' \cite{Gen59},
was suggested by de Gennes in 1959 on the basis of sum-rule arguments.
It implies that plane-wave density fluctuations with $q$ in the vicinity
of $q_{\rm max}$ decay slower in time due to the strong spatial correlations
that exist at these wave vectors.
Experimentally, this effect has been observed by coherent quasielastic 
neutron scattering on simple atomic fluids like noble gases \cite{noble}
and also on liquid metals \cite{metal}.
It has been confirmed by computer simulations on simple fluid
models\cite{BaZo94}.

The extension of this simple concept to molecular liquids
in the supercooled, i.\ e.\, high-density state,
is difficult and far less investigated \cite{MeKF87,ToWS98,ArRC96}.
Experimentally, only a few studies of the wave vector dependence of the
\arx\ in supercooled liquids \cite{MeKF87,ToWS98} indicate that a
narrowing is present.
The effect, however, was small and the analysis remained only 
{\em qualitative}.
Moreover, in glass forming polymers no evidence for any clear modulation
was reported in an extended analysis where all relaxation processes 
were taken into account \cite{ArRC96}.
The difficulties mainly arise from the following reasons:
(i) in the supercooled regime, the presence of memory effects leads 
to a structural \arx\ with a marked stretching of the correlation function;
(ii) for {\em molecular} liquids the intermolecular structure factor 
$S_{\rm m}(q)$ is a linear combination of many partial atom-atom 
structure factors;
(iii) the dynamical observables depend in general
upon both translational {\em and} 
rotational degrees of freedom.

In this letter we present the first {\em quantitative} 
coherent neutron scattering study of two very similar
substituted aromatic liquids Toluene and m--Toluidine.
A large $q$-range was covered including the main 
peak of the static structure factor $S_{m}(q)$ whose maximum is at 
$q_{\rm max}=1.8$\,\AA$^{-1}$.

In both liquids we find an extremely pronounced "de Gennes narrowing", 
not yet reported for the \arx\ of any glass forming liquid.
It is centered around a wave vector $q_0 < q_{\rm max}$, where $q_0$ 
is equal to the first maximum of the 
{\em center--of--mass static structure factor} $S_{\rm COM}(q)$.
Moreover, the wave-vector dependence of the mean structural relaxation 
time $\langle \tau_q \rangle$ can be {\it quantitatively} described by
$\langle \tau_q \rangle \propto S_{\rm COM}(q)/q^2$.
This indicates that around these wave-vectors the translational motion 
dominates the relaxation of $\Phi(q,t)$.

Fully deuterated Toluene (C$_7$D$_8$, T$_g$=117\,K, T$_m$=178\,K) 
was obtained from 
Euriso-top and sealed into quartz capillaries (Fa.~Hilgenberg, Malzfeld).
Optical inspection showed no signs of crystallization when cooling down 
to $T_g$.
Fully deuterated m--Toluidine (C$_7$D$_8$ND$_2$, $T_g$=183.5\,K, 
$T_m=243.9$\,K) 
was filled into a thin hollow aluminium cylinder.
Note that for temperatures around the calorimetric glass transition
the time scale of the  methyl group rotation
around their $C_3$--axis is several orders of magnitude faster than the
relevant $\alpha$-relaxation \cite{Hinze}.
Moreover, the quasielastic contribution of the rotational motion 
is negligible \cite{WiFW99}.
Both liquids are simple molecular glass formers and
particularly interesting as they can be thus considered as rigid planar
and small organic molecules.
Their dynamic has already been studied by a large variety
of experimental methods, e.\ g.\ viscosimetry, dielectric relaxation,
nuclear magnetic resonance, light- and incoherent neutron scattering
\cite{all}.

The main experiments were performed on the neutron spin echo spectrometer 
(NSE) IN11 of the Institut Laue Langevin in Grenoble, France.
IN11 was operated in a double spin echo configuration
with the new multidetector covering 30$^\circ$.
Different mean incident wavelengths $\bar \lambda_i$ and scattering angles
$2 \theta$ were chosen to cover the $q$ region around the first
diffraction peak \cite{MoDP97,MoAB98}.
The polarization was measured for several temperatures. 
Normalization was achieved by division by
the resolution function measured on a quartz rod of similar
shape and by the polarization at $t=0$.
We thus obtain the normalized density correlation
function $\Phi(q,t)=S(q,t)/S(q,0)$.
Additional experiments have been performed for Toluene on
the time-of-flight spectrometer (TOF) Mib{\'e}mol of the Laboratoire
L{\'e}on Brillouin in Saclay, France.
The dynamic structure factor $S(q,\omega)$ was measured 
with an elastic energy resolution $\Delta E$=36\,$\mu$eV giving
access to a $q$ range for elastic scattering of 
$0.2 \le q_{\rm el}\le 1.7$\,\AA$^{-1}$.
After standard corrections the data were interpolated to constant
$q$ with step $\Delta q=0.05$\,\A-1 to yield the dynamic structure factor
$S(q,\omega)$.

Spin-echo data are interpreted by using the stretched exponential function
of Kohlrausch
\begin{equation} \label{kww}
\Phi(q,t;T)=f_q \exp(-(t/\tau_q)^{\beta_q}), \quad 0< \beta_q <1
\end{equation}
where in principle all parameters can be temperature and $q$ dependent.
As $\langle \tau_q \rangle$ and $\beta_q$ are strongly correlated, we
use the mean $\alpha$-relaxation time
$\langle \tau_q \rangle = \int_{0}^{\infty} {\rm d}t \Phi(q,t)/f_q
=\beta_q^{-1}\Gamma(\beta_q^{-1})\langle \tau_q \rangle$
as a fitting parameter instead.

The individual correlation functions $\Phi(q,t;T)$ shown in Fig.~1
indicate already without any fitting that the relaxation is slowed 
down by about a factor of ten around $q\simeq1.3$\,\A-1 when compared 
to other $q$.
Due to the stretched character and the strong $q$--dependence of
$\langle \tau_q \rangle$ a much 
larger dynamic time window would be needed in order 
to cover all $q$--dependent $\Phi(q,t;T)$ at one given temperature.
Although a time range of about 2.5 decades is covered
our data do not really allow for reliable three parameter fits 
to the individual $\Phi(q,t;T)$.

However, quantitative information and consistent results with acceptable 
error bars can be extracted by combining measurements at different 
temperatures.
Without altering the interpretation of the underlying physics 
we apply the well-known time-temperature superposition principle
which means that the correlation functions $\Phi(q,t;T)$ can be 
superimposed simply by rescaling the time axis.
Its operational validity at least in a narrow temperature range of 
about $30$\,K in the time range of neutron scattering was demonstrated 
for a large number of glass forming substances 
\cite{MeKF87,ToWS98,RiFF88,WuPP96}.
We use a compilation of $\alpha$-relaxation times $\tal$
obtained by other spectroscopic techniques \cite{all}
to rescale the experimental times according to
$\hat t=t {\tal(T_0)} / {\tal(T)}$,
with an arbitrary normalization at $T_0=140$\,K and $T_0=230$\,K
for Toluene and m--Toluidine, respectively.
Indeed, for both liquids all $\Phi(q,t;T)$ collaps
onto a temperature--independent master curve $\Phi(q,\hat t)$
without any adjustment of $f_q(T)$ as illustrated in 
Fig.~2.
The successful scaling implies that the microscopic relaxation time 
measured by neutron scattering is proportional to the "macroscopic" 
relaxation time and that the exponents $\beta_q$ and the amplitudes 
$f_q$ cannot depend strongly on temperature.

>From a fit with (\ref{kww}) to the master curves $\Phi(q,\hat t)$ we 
obtain the relaxation time $\langle \tau_q(T_0) \rangle$ which are
displayed in Fig.~3 normalized to its maximum value at $q_0$.
For both liquids, the relaxation time $\langle \tau_q \rangle$ has an 
overall variation with $q$ (far from the hydrodynamic regime on the 
low-$q$ side and from the free-particle behavior on the high-$q$ side)
of a factor of ten.
It passes through a maximum as $q$ is varied from 
$0.8$ to $2.0$\,\AA$^{-1}$. 
To our knowledge this variation is much stronger than that in any other
liquid studied so far. 
More strikingly, the strength of the variation of $\langle \tau_q \rangle$ 
with $q$ and the position of its maximum can {\em quantitatively} be 
described by $\langle \tau_q \rangle \propto S_{\rm COM}(q)/q^2$
where the center--of--mass static structure factor $S_{\rm COM}(q)$ was 
obtained by Monte-Carlo simulations \cite{MoDP97,MoAB98}.
It was not expected that de Gennes's very general concept \cite{Gen59} 
would strickly apply to molecular liquids like Toluene or m--Toluidine.
It is far from being trivial because the correlation functions are stretched 
and the relaxation times are long compared to those of atomic fluids.
In fact, our findings indicate that in the $q$-range corresponding to 
molecular length scales (van der Waals radius)
and in the $\alpha$--relaxation regime, neutron scattering mainly 
probes the motion of the centers of mass, 
i.e. mainly probes the translational motion of the molecules.
This result is of fundamental importance for the interpretation of 
neutron scattering spectra in general.
What we actually measure are, of course, atom-atom correlations 
involving the H,D and C nuclei.
In addition to a translational component, these correlations also contain
contributions involving the rotational motions.
Our joint use of Monte--Carlo simulations and neutron scattering experiments
indicates that at the $q$ values considered
and for the $\alpha$--relaxation, these latter contributions are
smallcompared to the translational one.
This finding is also of particular interest for microscopic glass 
transition theories like the mode--coupling (MC) theory, either
in its idealized form \cite{Got91} or in the molecular version \cite{mol};
indeed in MC theory the static structure factor is the only major input
that determines
the slowing down of the \arx\. The
MC theory gives at least a qualitative description of the q-dependence of 
$\langle \tau_q \rangle$ around the static structure factor maximum 
\cite{FuHL92} and the importance of the center--of--mass slow dynamics 
in this framework is illustrated by the so-called semi schematic 
model \cite{FaSS98}.


To underline the robustness of our NSE results $\langle \tau_q \rangle$ has 
been independently obtained for Toluene from Mib{\'e}mol experiments by 
fitting (\ref{kww}) the $S(q,\omega)$ at 200\,K in frequency space convoluted 
with the measured resolution function (see Fig.~3).
The stretching $\beta_q$ was kept fixed to 0.5 close to the average
value 0.48 obtained from our spin--echo experiment (see below).
As can be judged from Fig.~3, excellent agreement within a few percent 
between NSE and TOF is found for the $q$--dependence of the relaxation time
over a large $q$ range.
To emphasis this good agreement even further we included in Fig.~2 
$\Phi(q,\hat t)$ obtained from Mib{\'e}mol spectra $S(q,\omega)$ by 
Fourier deconvolution with the measured resolution function. 
Thus, an identical wave--vector dependence of $\langle \tau_q \rangle$ 
is observed from the stable liquid state, where the typical 
relaxation times are of the order of picoseconds, to the
supercooled state, with relaxation times of the order of nanoseconds.
Such a behavior was also observed in orthoterphenyl (OTP) \cite{ToWS98}.

Another important output of our study 
is the $q$--dependence of the amplitude $f_q$ of the $\alpha$--relaxation. 
In Fig.~4,
$f_q$ shows a maximum at $q=1.4$\,\AA$^{-1}$ which is very close to the 
maximum of the center--of--mass static structure factor $S_{\rm COM}(q)$, 
in agreement with the
numerical solutions of mode--coupling equations \cite{mol}. 
However, contrary to this latter,
it does not oscillate in phase with $S_{\rm COM}(q)$ (nor $S_{m}(q)$),
and only a very smoothly decreasing behavior is observed at higher $q$ 
due to additional rotational contributions\cite{Teboul}.
We could not detect any systematic $q$--dependence of the stretching parameter 
$\beta_q$ which is at variance to theoretical predictions \cite{FuHL92}
and to the results in OTP \cite{ToWS98}.
Note however, that in OTP no variation of $\beta_q$ with $q$ was
detected at wave numbers corresponding to the center--of--mass correlations
but at larger $q$ values.
>From NSE we find an average $\langle \beta_q \rangle_q \simeq 0.48 \pm 0.02$
for Toluene and $\langle \beta_q \rangle_q \simeq 0.58 \pm 0.02$ for 
m--Toluidine.
The conjecture of $q$ dependent oscillations of $\beta_q$ and
$\langle \tau_q \rangle$ indicates a systematic error source in 
NSE experiments:
since we have used a wave vector band 
$\Delta q /q = \Delta \lambda /\lambda \simeq 10$\,\%
which is nearly as broad as the diffraction peak itself \cite{MoDP97,MoAB98}
we have actually measured averages $\langle \Phi(q,t;T) \rangle_q$
of correlation functions with possibly different parameters $\beta_q$ 
and $\langle \tau_q \rangle$.
In practice, a sum of different stretched exponentials can again be
fitted by a Kohlrausch with a smaller effective exponent 
$\beta_{\rm eff}$ and a different $\langle \tau_q \rangle_{\rm eff}$.
>From numerical examples using 
$\langle \tau_q \rangle \propto q^{-2}S_{\rm COM}$
and $0.4 \le \beta_q \le 0.5$ as input we conclude that the difference between 
$\langle \tau_q \rangle_{\rm eff}$ and $\langle \tau_q \rangle$ is always 
less than 10\,\% and the difference in the line-shape is anything between 
0.01 and 0.05 which is far less than our experimental accuracy {\it i.\ e.} 
our conclusion remain valid.

The main result of our communication concerns the wave vector dependence 
of the $\alpha$ relaxation time $\langle \tau_q \rangle$.
The maximum in $\langle \tau_q \rangle$ at $q_0< q_{\rm max}$ can be 
interpreted as a de Gennes narrowing for molecular 
glass forming liquids resulting from the strong spatial correlations 
between the centers--of--mass of the molecules.
The relaxation time characteristic is quantitatively described by 
$\tau_q \propto S_{\rm COM}(q)/q^2$.
$S_{\rm COM}(q)$ is the spatial Fourier transform of the orientationally 
averaged pair correlation function.
On molecular scales and microscopic times, the time evolution of density 
fluctuations of the glassforming Toluene and m--Toluidine is dominated by 
translational dynamics, {\it i.\ e.} the center--of--mass 
relaxational dynamics. 
Such a strong de Gennes narrowing effect has not yet been reported for 
the $\alpha$-relaxation of low molecular-weight glass-forming liquids.
Whether $S_{\rm COM}(q)$ is the right input for a sophisticated mode-coupling
analysis in the supercooled high-density state can only be decided by
further molecular dynamics simulations \cite{Teboul}.
Some fortuitous compensation effects in the orientational fluctuations 
can play an important role.

Despite the structural similarity of both molecules there is an
important difference that might play an interesting role in the
structural relaxation dynamics: the formation of hydrogen bonds 
in a system like m--Toluidine induces clustering of a few molecules 
whose signature is a well defined prepeak around 
$q\simeq0.5$\,\AA$^{-1}$ in the static structure factor \cite{MoAB98}.
Unfortunately, our experiments in this low $q$-range are of too low 
statistical qualitiy --- NSE experiments outside the diffraction maximum are 
extremely difficult to perfom --- to decide whether this static heterogeneity
has a counterpart in the dynamics or not.

\end{multicols}

\newpage

{\bf Figure captions} \\

Figure 1: \\
Density correlation functions $\Phi(q,t)$ at 160\,K for
Toluene at three different wave vectors $q$.
Lines are fits with the Kohlrausch function (\ref{kww}).
Note, at $q_0=1.3$\,\A-1, close to the first maximum of the 
center--of--mass static structure factor $S_{\rm COM}(q)$, 
the correlations decay much slower than for other wave vectors.
 
Figure 2: \\
Master curves $\Phi(q,\hat t)$ at several $q$ for Toluene (a)
and m--Toluidine (b) as obtained by rescaling times to
$\hat t=t \tal(T_0)/\tal(T)$.
The data fall onto a temperature-independent master curve.
The dense points at long times steem from Fourier 
deconvoluted Mib{\'e}mol TOF spectra $S(q,\omega)$.
For Toluene $\Phi(q,\hat t)$ at 200\,K 
for $q=0.9$\,\A-1 and $q=1.3$\,\A-1\ is included (a). 
For m--Toluidine $\Phi(q,\hat t)$ for $q=1.3$\,\A-1\ from a fragmentary 
experiment on Mib{\'e}mol at 273\,K is included in (b).
Note the excellent agreement between NSE and TOF.

Figure 3: \\
(a) Wave vector dependence of the structural relaxation time 
$\langle \tau_q(T) \rangle / \langle \tau_{q_0}(T) \rangle$ for Toluene
obtained from fits with (\ref{kww}) to the neutron spin--echo master 
curves (cf.\ Fig.~2) (open symbols). Results from fits with the Fourier 
transformation of (\ref{kww}) to the dynamic structure factor 
$S(q,\omega)$ from Mib{\'e}mol at $200$\,K are included (closed symbols).
The solid line is $q^{-2} S_{\rm COM}(q)$. 
obtained by Monte-Carlo simulations at 160\,K.
The dotted lines is the experimentally observed total
$q^{-2} S_{\rm m}(q)$ at 161\,K \cite{MoDP97,MoAB98}.
(b) The same for m--Toluidine with $q^{-2} S_{\rm COM}(q)$ at 260\,K
and $q^{-2} S_{\rm m}(q)$ at 248\,K.

Figure~4: \\
Wave vector dependences of the the prefactor $f_q$ obtained from fits
of (\ref{kww}) to the master curves.
Open squares: Toluene (NSE), open circles: m--Toluidine (NSE).
The closed squares are plateau values from Mib{\'emol} TOF data
obtained by time averaging $\Phi(q,t)$ at 140\,K in the range 
3\,psec $\le t \le$ 8\,psec and multiplied by 1.1 to match NSE data.
Solid line: $S_{\rm COM}(q)$, dotted line $S_{\rm m}(q)$ for Toluene
as in Fig.~3.

\end{document}